\documentclass[
	aps,
	prl,
	amsmath,
	reprint,
	superscriptaddress,
	longbibliography
]{revtex4-2}
\usepackage{hyperref}
\usepackage{graphicx}
\usepackage{units}
\usepackage{upgreek}
\usepackage{color}
\usepackage{amsmath, amssymb, amsfonts,enumerate}
\makeatletter
\renewcommand\subparagraph{\@startsection{subparagraph}{5}{\z@}{3.25ex\@plus1ex\@minus.2ex}{-1em}{\normalfont\normalsize\bfseries}}

\renewcommand{\@make@capt@title}[2]{\@ifx@empty\float@link{\@firstofone}{\expandafter\href\expandafter{\float@link}}{\textsf{\bfseries#1}}\@caption@fignum@sep{\textsf{#2}}}

\renewcommand\@caption@fignum@sep{\textsf{\bfseries.}}

\let\oldtheequation\theequation
\renewcommand\tagform@[1]{\maketag@@@{\ignorespaces#1\unskip\@@italiccorr}}
\renewcommand\theequation{(\oldtheequation)}
\makeatother

\usepackage{tabularx}

\begin{document}

\begin{abstract}
	{\textbf{Abstract}}: There is currently a strong interest in the collective behavior of chiral active particles that can propel and rotate themselves. In the presence of alignment interactions for many chiral particles, chiral self-propulsion can induce vortex patterns in the velocity fields. However, these emerging patterns are non-permanent, and do not induce global vorticity. Here we combine theoretical arguments and computer simulations to predict a so-far unknown class of collective behavior. We show that, for chiral active particles, vortices with significant dynamical coherence emerge spontaneously. They originate from the interplay between attraction interactions and chirality in the absence of alignment interactions. Depending on parameters, the vortices can either feature a constant vorticity or a vorticity that oscillates periodically in time, resulting in self-reverting vortices. Our results may guide future experiments to realize customized collective phenomena such as spontaneously rotating gears and patterns with a self-reverting order.
\end{abstract}

\title{Self-reverting vortices in chiral active matter}

\author{L. Caprini}
\email{lorenzo.caprini@hhu.de, lorenzo.caprini@uniroma1.it}
\affiliation{Heinrich-Heine-Universit\"{a}t D\"{u}sseldorf, Universit\"atsstrasse 1, D-40225 D\"{u}sseldorf, Germany.}
\affiliation{University of Rome La Sapienza, Piazzale Aldo Moro 5, 00185  Rome, Italy.}

\author{B. Liebchen}
\affiliation{
Technische Universität Darmstadt, Hochschulstrasse 8, 64289 Darmstadt, Germany.}

\author{H. L\"owen}
\affiliation{Heinrich-Heine-Universit\"{a}t D\"{u}sseldorf, Universit\"atsstrasse 1, D-40225 D\"{u}sseldorf, Germany.}

\date{\today}

\maketitle

\section{Introduction}
Chirality refers to the property of objects to be non-superimposable on their mirror images.
The concept originated in the mid-19th century and it is attributed to the chemist Louis Pasteur, who observed that crystals of tartaric acid exist in two distinct, non-superimposable forms, which he referred to as "right-handed" and "left-handed."
Also more than a century ago, Bronn, Jennings, and others realized that shape-asymmetric motile microorganisms generically follow chiral trajectories~\cite{bronn1862klassen,jennings1901significance}, i.e.\ they do not only self-propel but they also self-rotate, showing circular trajectories.
Recently, the discovery of synthetic colloidal microswimmers in the 21st century~\cite{ismagilov2002autonomous, paxton2004catalytic, dreyfus2005microscopic} has stimulated a significant interest in chiral self-propelled particles~\cite{lowen2016chirality, liebchen2022chiral}:
Like their biological counterparts, these particles also generically follow circular trajectories if they feature a shape-anisotropy~\cite{kummel2013circular, denk2016active, campbell2017helical,workamp2018symmetry, liu2019collective, vega2022diffusive, siebers2023exploiting} or are torqued by an external field~\cite{massana2021arrested}. 
In addition, it is now known that chirality can emerge due to
hydrodynamic interactions with walls or interfaces, as in bacteria~\cite{lauga2006swimming, petroff2015fast, perez2019bacteria}, or
due to memory effects in viscoelastic environments and droplet swimmers~\cite{narinder2018memory, carenza2019rotation,feng2023self}.

\begin{figure*}[!t]
\centering
\includegraphics[width=1\linewidth,keepaspectratio]{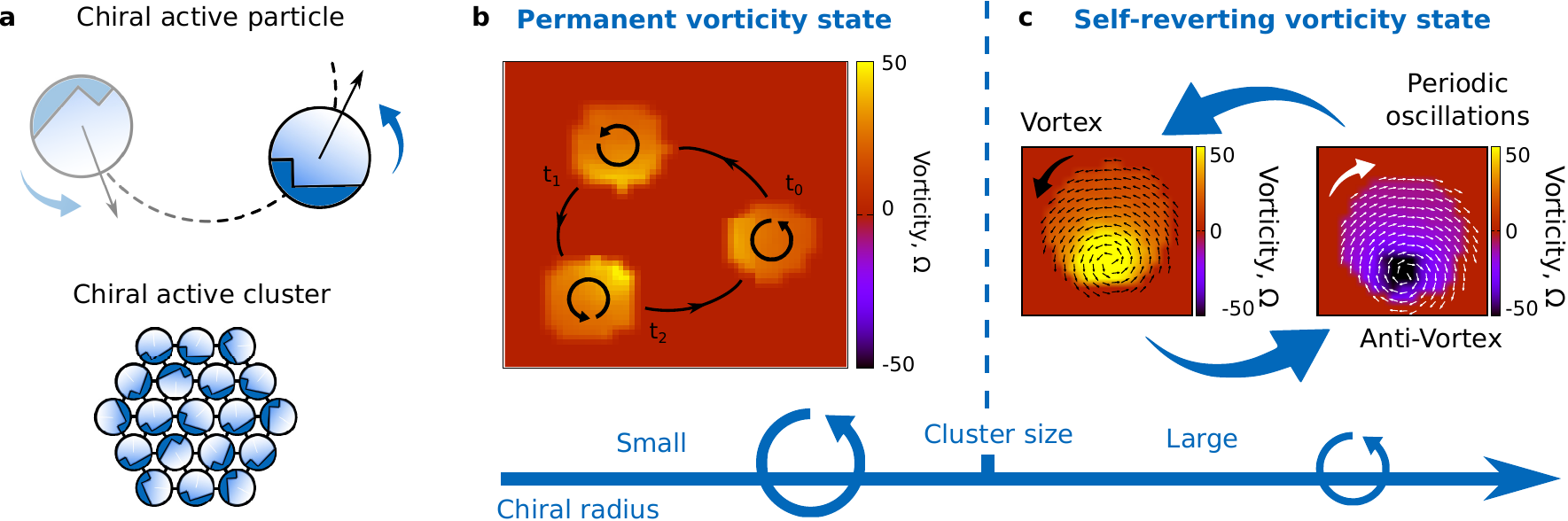}
\caption{ \textbf{Chirality-induced collective motion.}
(a): Illustrations of chiral active particles. The particle orientation is indicated by a dark blue cap with L-shape responsible for the particle chirality.
The black and blue arrows serve as a schematic representation of particle velocity and chirality. We sketch the typical rotating trajectory of a chiral active particle and illustrate a chiral active cluster maintained by attractive interactions.
(b): Time series of snapshot configurations showing the permanent vorticity state, as revealed by the vorticity field (colors). The cluster exhibits collective rotations, following circular trajectories and displaying counterclockwise vorticity.
(c): Time series of snapshots for the cluster exhibiting spinning dynamics and revealing the self-reverting vorticity state, i.e. a periodic alternation between vortex and antivortex configurations.
Arrows indicate the velocity field while colors denote the vorticity value.
(b)-(c) are obtained with cluster size square $L_c^2=904$ and reduced chirality $\omega \tau=10, 10^2$, respectively.
The remaining parameters of the simulations are: $\tau_I/\tau=10^{-6}$, $\text{Pe}=\tau v_0/\sigma=50$, $\tau^2 \epsilon/(\sigma^2 m)=5\times 10^{3}$, $\tau^2 T /(m\sigma^2)=10^{-5}$.
}\label{fig:Fig1}
\end{figure*}
At the many-particle level in the presence of alignment interactions, chiral self-propulsion can induce a rich panorama of collective phenomena such as pattern formation~\cite{liebchen2017collective, levis2018micro, ai2018mixing, kruk2020traveling, liao2021emergent, negi2022geometry, zhang2022collective, kreienkamp2022clustering, hiraiwa2022collision, lei2023collective, negi2023geometry, ceron2023diverse}, which includes rotating micro-flock patterns~\cite{liebchen2017collective}, 
chiral self-recognition~\cite{arora2021emergent}, traveling waves~\cite{liebchen2016pattern}, and even chimera states~\cite{kruk2020traveling}.
By contrast, in the absence of alignment, circular motion does not generally show exciting collective effects. Indeed, in repulsively interacting chiral active particles, it primarily contributes to reducing obstacle accumulation~\cite{caprini2019activechiral, fazli2021active, caprini2023chiral}. Consequently, it suppresses the clustering typical of repulsive active particles~\cite{liao2018clustering, ma2022dynamical, semwal2022macro, sese2022microscopic, bickmann2022analytical}, by contrast, leading to a hyperuniform phase~\cite{lei2019nonequilibrium, huang2021circular, zhang2022hyperuniform, Kuroda_2023} and demixing~\cite{reichhardt2019reversibility, ai2023spontaneous}. Indeed, chirality only affects the particle's ability to continuously explore space, reducing the effective persistence of the activity.
However, recently, Debets et al. have discovered a peculiar oscillatory caging effect in chiral active glasses~\cite{debets2023glassy} while Liao and Klapp have observed intriguing vortex patterns in the velocity fields for significant chirality levels~\cite{liao2018clustering}. Despite the relevance of the latter study, these structures are non-permanent and do not induce global vorticity as also observed in chiral rollers~\cite{zhang2020reconfigurable}.

In the present work, we combine theoretical arguments and particle-based simulations to predict the existence of a so-far unknown class of structures in chiral active matter. First, and perhaps least surprisingly, for low chirality (low self-rotation frequency) we find that attractive chiral active particles (Fig.~\ref{fig:Fig1}~a) without alignment interactions form moving rigid clusters that feature full velocity-alignment of the contained particles~\cite{caprini2023flocking} and spatial velocity correlations~\cite{caprini2020spontaneous, caprini2020hidden} but vanishing vorticity. 
However, for high chirality, we observe a transition to a rotation pattern that is characterized by a persistent and time-independent vorticity (Fig.~\ref{fig:Fig1}~b) and is termed {\it permanent vortex state}.
This state can be viewed as the superposition of the translational motion characterizing the previous state and an additional collective rotation due to chirality, which transfers from the single particle to the collective level.
For even higher chirality the rotation pattern again changes and the vortex starts to dynamically revert itself, exhibiting periodic transitions between vortex and antivortex configurations (Fig.~\ref{fig:Fig1}~c). We refer to these structures as {\it self-reverting vortices}.
The occurrence of this state is a consequence of the competition between chirality and isotropic interactions, which suppresses the tendency of a cluster to collectively rotate.

\section{Results}
\subsection{Model for chiral particles}
To concretely investigate these states, we consider a system of $N$ interacting active chiral particles with mass $m$, where each particle is governed by underdamped equations of motion for their positions, $\mathbf{x}_i$, and velocities, $\mathbf{v}_i = \dot{\mathbf{x}}_i$. Every particle is in contact with a thermal bath at temperature $T$ and experiences a frictional force, $\gamma\mathbf{v}_i$, with friction coefficient $\gamma$. 
Activity is incorporated in the dynamics as a stochastic force, which imparts to each particle a constant swim velocity, $v_0$, along with an orientation vector, $\mathbf{n}_i=(\cos{\theta_i}, \sin{\theta_i})$. 
Here, $\theta_i$ are the orientational angles and, in accordance with the active Brownian particle (ABP) model~\cite{shaebani2020computational} describing circular swimmers~\cite{van2008dynamics, sevilla2016diffusion, reichhardt2019active, chepizhko2020random, han2021fluctuating, van2022role, olsen2021diffusion}, evolves as Brownian noise with a constant drift angular velocity, $\omega$. The latter is also known as particle chirality and is responsible for circular trajectories~\cite{van2008dynamics}.
Thus, the system's dynamics can be expressed as
\begin{subequations}
\label{eq:wholeABPdynamics}
\begin{align}
\label{eq:v_dynamics}
m\dot{\mathbf{v}}_i &=-\gamma{\mathbf{v}}_i + \mathbf{F}_i + \gamma v_0 \mathbf{n}_i + \sqrt{2T \gamma}\boldsymbol{\eta}_i \\
\label{eq:theta_dynamics}
\dot{\theta}_i&= \sqrt{2D_r} \xi_i + \omega \,.
\end{align}
\end{subequations}
Here, $D_r$ represents the rotational diffusion coefficient, and $\xi_i$ and $\boldsymbol{\eta}_i$ denote white noises with zero average and unit variance. The particle chirality $\omega$ determines the characteristic radius of the circular trajectory displayed by a single active chiral particle, specifically $v_0/\omega$. 
In this system, the absence of torques between particles results in their sole interaction through the force $\mathbf{F}_i = -\nabla_i U_{tot}$, where $U_{tot} = \sum_{i<j} U(|{\mathbf r}_{ij}|)$, and ${\mathbf r}_{ij} = \mathbf{x}_i -\mathbf{x}_j$.
The shape of the interaction potential $U(r)$ is obtained by truncating and shifting an attractive Lennard-Jones potential, $U_{LJ}(r) = 4\epsilon\left[\left(\frac{\sigma}{r}\right)^{12} - \left(\frac{\sigma}{r}\right)^{6}\right]$. 
The potential $U(r)$ is therefore defined as $U(r)=U_{LJ}(r)- U_{LJ}(3\sigma)$ for $r\leq 3\sigma$, and zero otherwise.
Here, $\sigma$ signifies the nominal particle diameter, while $\epsilon$ stands for the energy scale of the interactions.
The interparticle attraction is sufficiently large to guarantee that the cluster structure remains stable.
The system is characterized by three primary time scales: the inertial time, $\tau_I = m/\gamma$, determining the velocity relaxation; the persistence time, $\tau = 1/D_r$, which dictates the duration required for active particles to randomize their orientations; 
and the time $1/\omega$ necessary for the orientation to complete a full rotation due to chirality.
We remark that our model considers self-propulsion and chirality, i.e.\ self-rotations, as two independent mechanisms.
Indeed, even if these two propulsions are often related in chiral active colloids, this is not the case in other physical systems, for instance active granular particles with an intrinsic chirality and spinners where the self-propulsion is even absent.

\begin{figure*}[!t]
\centering
\includegraphics[width=1\linewidth,keepaspectratio]{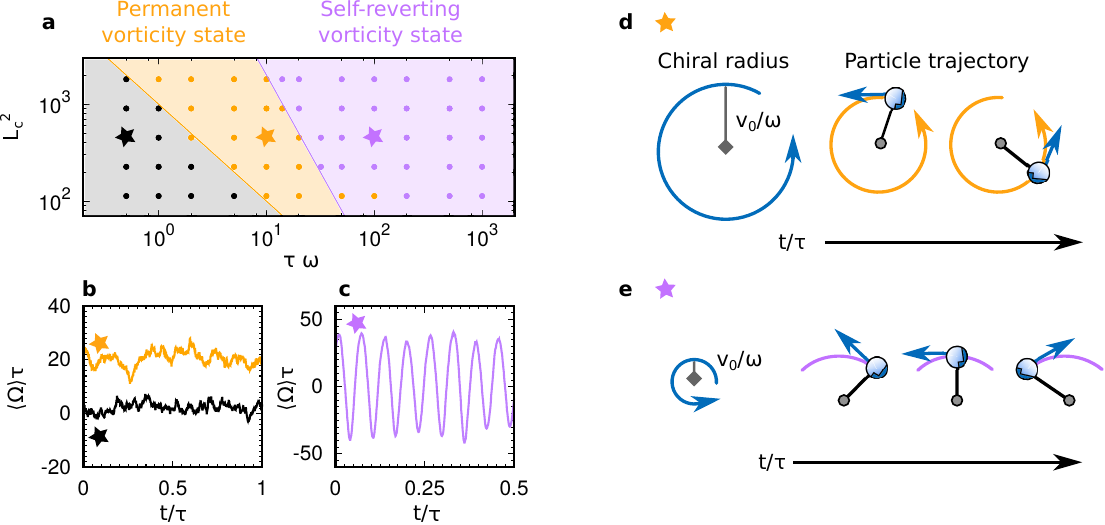}
\caption{ \textbf{Vorticity states.}
(a): State diagram for chiral active particles in the plane of reduced chirality $\omega\tau$ and cluster size square $L_c^2$.
Black, orange, and purple colors denote negligible vorticity, permanent vorticity, and self-reverting vorticity states, respectively.
(b)-(c): Spatial average chirality, $ \langle \Omega \rangle\tau$, as a function of the rescaled time $t/\tau$. The three time-trajectories in (b) and (c) correspond to the stars in (a): Specifically, the black, orange, and purple curves are obtained for $\omega\tau=5\times 10^{-1}, 10, 10^2$ with $L_c^2=452$. The remaining parameters of the simulations are: $\tau_I/\tau=10^{-6}$, $\text{Pe}=\tau v_0/\sigma=50$, $\tau^2 \epsilon/(\sigma^2 m)=5\times 10^{3}$, and $\tau^2 T /(m\sigma^2)=10^{-5}$.
(d)-(e): Illustrations of a chiral active dumbbell, anchored to one of the particles. The permanent vorticity state occurs when the chiral radius $v_0/\omega$ (grey line centered on a diamond) is larger than the dumbbell size (black line centered on a circle), allowing the mobile particle to complete a full rotation around the other (orange star). The self-reverting vorticity state occurs in the opposite regime, such that the mobile particle cannot complete a full rotation around the other. Indeed, its self-propulsion is reversed before completion resulting in partial clockwise and counterclockwise rotations (purple star).
}\label{fig:Fig2}
\end{figure*}

\subsection{Theoretical prediction for vortex states}
To understand the collective behavior of attractive chiral active particles, we first develop a mapping showing that the overall dynamics are governed by the competition of velocity alignment and an effective Lorentz force.
Specifically, Eqs.~\eqref{eq:wholeABPdynamics} can be mapped to alternative dynamics using an exact change of variables and a lattice approximation, applicable to strongly attractive active particles in the large persistence time regime (see Methods).
The evolution of the particle velocity $\mathbf{v}_i$ is effectively described by:
\begin{equation}
\label{eq:dynamics_explanation}
\dot{\mathbf{v}}_i = \frac{1}{\gamma}\sum_j \mathbf{J}_j \cdot\left(\mathbf{v}_i - \mathbf{v}_j\right) +\omega \mathbf{v}_i \times \mathbf{z} \,.
\end{equation}
Here, $\sum_j$ encompasses neighboring particles to the $i$-th one, $\mathbf{z}$ represents the normal vector to the plane of motion, and the elements of the matrix $\mathbf{J}_j$ are reported in the Methods.
Equation~\eqref{eq:dynamics_explanation} holds in the large persistence regime and reveals that the system's behavior is primarily governed by two distinct forces.
The first one, independent of chirality, manifests as an effective alignment force emerging from the interplay between interactions and activity. It accounts for the observed velocity alignment and, indeed, is minimized in three particle configurations: i) full alignment; ii) vortex; iii) antivortex (see Methods).
In the absence of chirality, there is no preference among these configurations.
The second term in Eq.~\eqref{eq:dynamics_explanation} is solely induced by chirality, being $\propto \omega$, and operates as an effective Lorentz force.
This indicates that chirality influences the dynamics of an active particle, akin to an effective magnetic field, responsible for particle rotations.
When $\omega$ is comparable to $\mathbf{J}_j$, the effective magnetic field selectively promotes vortex or antivortex states for negative and positive $\omega$, respectively. Consequently, this analytical argument predicts the spontaneous emergence of substantial vorticity in the system.

\subsection{Simulations unveil self-reverting vortices}
To observe the predicted vortex-states, we perform simulations within a box of size $L$ under periodic boundary conditions, ensuring that the particle packing fraction $\phi=(N/L^2) \sigma^2 \,\pi/4=0.3$ remains constant. 
It is crucial to emphasize that our findings pertain to the large persistence regime ($\tau/\tau_I\gg1$), resulting in effectively overdamped dynamics (see Methods). Given this choice, the same results can be obtained by considering an overdamped dynamics if the thermal noise is sufficiently small.
The condition $\tau/\tau_I\gg1$ signifies that the persistence length $v_0\tau$ is the dominant length scale in the system, notably larger than the cluster size $L_c \approx \sigma \sqrt{N}_c$, where $N_c$ is the number of particles in the cluster. 
We remark that in the opposite small persistence time regime $\tau/\tau_I\ll1$, the system is close to equilibrium and the active force behaves as thermal noise (see Methods). Thus, no collective motion can be observed in this regime.
Therefore, we conduct a numerical study by keeping fixed $\tau/\tau_I\gg1$.
In addition, the dynamical states shown here are obtained only in the regime of large attractions compared to thermal noise strength and activity, i.e.\ when the typical potential energy due to the interparticle interactions is large compared to the thermal energy and the kinetic energy associated with self-propulsion $\approx m v_0^2/2$.
Indeed, without this condition, the cluster is not stable because particles are able to leave it and therefore it is not possible to observe collective motion.
Here, to investigate the influence of chirality, we vary the associated dimensionless parameter, the reduced chirality $\omega\tau$, and examine different cluster sizes $L_c$.

\begin{figure*}[!t]
\centering
\includegraphics[width=0.95\linewidth,keepaspectratio]{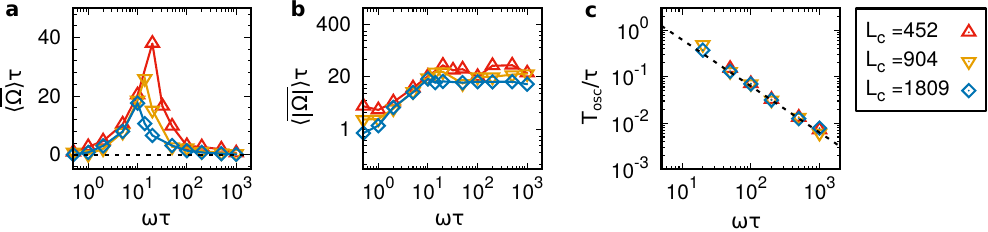}
\caption{ \textbf{Characterization of the vorticity states.}
	(a): Time-averaged value of the spatial average vorticity $\bar{\langle \Omega\rangle}\tau$, as a function of the reduced chirality $\omega\tau$.
	(b): Time-average of the modulus of the rescaled spatial average vorticity, $\bar{\langle |\Omega|\rangle}\tau$, as a function of $\omega\tau$.
	(c): Time-period of the vorticity oscillations in the self-reverting vorticity state as a function of $\omega\tau$. 
The dashed black line plots the function $2\pi/(\omega\tau)$.
	The analysis in (a)-(c) is performed for different values of the cluster size square, $L_c^2=\sigma^2N_c$, with specific values provided in the external legend.
	The remaining parameters of the simulations are: $\tau_I/\tau=10^{-6}$, $\text{Pe}=\tau v_0/\sigma=50$, $\tau^2 \epsilon/(\sigma^2 m)=5\times 10^{3}$, $\tau^2 T /(m\sigma^2)=10^{-5}$.
}\label{fig:Fig3}
\end{figure*}

We discover phenomena in active systems uniquely induced by circular motion and attractive forces. 
Reduced chirality $\omega\tau$ fosters collective rotational motion, with the entire cluster tracing persistent circular trajectories (see Supplementary Movie 1).
For further increasing values of $\omega\tau$, the cluster displays spinning dynamics rather than a circular trajectory. 
Indeed, the center of mass of the cluster undergoes rotations with a characteristic radius smaller than the cluster size $L_c$ (see Supplementary Movie 2).
To characterize rotational motion, we monitor the evolution of the spatial average of the vorticity field $\langle \Omega \rangle$, defined as
\begin{equation}
\langle\Omega\rangle = \int d{\mathbf{r}} \left(\partial_x v_y - \partial_y v_x \right) \,,
\end{equation}
which reads zero for particle velocities aligned in the same direction but assumes positive and negative values for antivortex and vortex configurations, respectively. 
In the limit of vanishing chirality $\omega\tau$ (black curve in Fig.~\ref{fig:Fig2}~b), the time-trajectory of $\langle\Omega(t)\rangle$ fluctuates around zero, signifying the absence of a preferred vorticity. 
By contrast, as $\omega\tau$ increases, $\langle\Omega(t)\rangle$ exhibits minor time-fluctuations around a value greater than zero, indicating a positive spatial average vorticity aligned with the single-particle chirality (Fig.~\ref{fig:Fig2}~b).
The breaking of rotational symmetry of a single chiral particle propagates to the collective level, resulting in a non-zero global vorticity: These configurations are identified as permanent vorticity states.
Conversely, when the cluster exhibits spinning dynamics, $\langle\Omega(t)\rangle$ displays periodic-time oscillations (Fig.~\ref{fig:Fig2}~c). 
This implies that particle velocities periodically switch between vortex and antivortex configurations, i.e. the cluster exhibits a self-reverting vorticity.
This phenomenon is a consequence of the additional time scale introduced by chirality, as confirmed by the oscillation period which scales as $\sim 1/\omega$.

\subsection{Mechanism behind self-reverting vorticity}
Permanent vorticity and self-reverting vorticity states can be intuitively explained by considering a chiral particle anchored to a fixed point with size determined by the strong attraction, $\sigma$.
Chirality enables the self-propulsion force to persistently rotate at a frequency of $1/\omega$ and, consequently, induces circular motion in the particle around the fixed point.
If the radius of the circular trajectory is larger than the distance with the fixed point, $v_0/\omega > \sigma$ (low chirality), the particle performs complete, persistent rotations, in the direction promoted by chirality (Fig.~\ref{fig:Fig2}~d).
This simple mechanism generates the permanent vorticity state at the collective level.
By contrast, in the opposite regime of large chirality ($v_0/\omega < \sigma$), chirality completely reverses the direction of the active force before the particle completes a rotation of $\pi$ radians around the immobile particle.
Consequently, the particle moves backward and forward, effectively alternating between clockwise and counterclockwise rotations (Fig.~\ref{fig:Fig2}~e). This explains the observed self-reverting vorticity state at the collective level.
This idea can be further supported by calculating the total torque ${\bf M}$ acting on the cluster, which is dominated by the outer particle layer at distance $R$ from the middle of the cluster (see Methods)
\begin{equation}
\label{eq:4main}
\mathbf{M}\approx \gamma v_0 \sum_{i=1}^{N_c}   (  \mathbf{r}_i \times\mathbf{n}_i ) \delta(|\mathbf{r}_i| -R)\,.
\end{equation}
Here $\mathbf{r}_i$ is a vector pointing from the center of the cluster to the $i$-particle position.
After time $\pi/\omega$, each $\mathbf{n}_i$ rotates by $\pi$ and thus also its spatial average. 
However, if $\mathbf{n}_i$ rotates with a period $\pi/\omega$ smaller than the period of $\mathbf{r}$, ${\bf M}$ is continuously subject to sign changes before a full cluster rotation: the cluster displays self-reverting vorticity.
By contrast, in the opposite regime, ${\bf M}$ never changes sign and the system displays a permanent vorticity state.

\subsection{State diagram}
Our findings are systematically explored by varying the cluster size $L_c$ and the reduced chirality $\omega\tau$ on a state diagram (Fig.~\ref{fig:Fig2}~a). We identify different states with different colors: non-permanent vorticity states (black dots) when the average vorticity is smaller than its time fluctuations; permanent vorticity states (orange dots), when the previous condition is fulfilled; self-reverting vorticity state (purple dots) when vorticity displays periodic oscillations.
Consistent with our intuitive explanation, the transition line between permanent and self-reverting vorticity state occurs when the cluster size $L_c=\sigma \sqrt{N}_c$ approaches the typical radius of the chiral trajectory $\sim v_0/\omega$.
This argument suggests the following scaling law
\begin{equation}
\label{eq:scalinglaw}
N_c \sim \frac{v_0^2}{\sigma^2\omega^2} \,,
\end{equation}
which fairly reproduces our numerical results (Fig.~\ref{fig:Fig2}~a).
This scaling law suggests that a larger cluster size favors the self-reverting vorticity state over the permanent vorticity state. 
Additionally, it is worth noting that an increase in cluster size promotes permanent vorticity states over states without vorticity. This is because the time fluctuations of the vorticity field decrease with increasing $N_c$.
We remark that the crossover between different states is not a sharp transition but occurs smoothly. Indeed, the regions in Fig.~\ref{fig:Fig2}~a are obtained by following the threshold criterion defined in the methods section.

To quantitatively characterize the different states, we consider the time average of the spatial average vorticity as a order parameter
\begin{equation}
\bar{\langle \Omega\rangle} = \lim_{t\to\infty}\frac{1}{t}\int dt \langle\Omega(t)\rangle \,.
\end{equation}
This observable shows a non-monotonic behavior with the reduced chirality $\omega\tau$ for different values of the cluster size $L_c$ (Fig.~\ref{fig:Fig3}~a).
For vanishing $\omega\tau$, the absence of permanent vortices (black state in Fig.~\ref{fig:Fig2}~a) induces rather small values of $\bar{\langle \Omega\rangle}$.
The increase of $\omega\tau$ enhances the value of $\bar{\langle \Omega\rangle}$ until it becomes larger than its typical time fluctuations and the system approaches the permanent vorticity state.
In this regime, $\bar{\langle \Omega\rangle}$ monotonically increases until a maximum is achieved.
This maximum occurs before the system approaches the self-reverting vorticity state, for which the periodic oscillations sharply lead to vanishing values of $\bar{\langle \Omega\rangle}$.
The amplitude of these oscillations is investigated by evaluating the time average of the modulus of the spatial average vorticity, $\bar{\langle |\Omega|\rangle}$ (Fig.~\ref{fig:Fig3}~b).
This observable monotonically increases with $\omega\tau$ until the self-reverting vorticity state is approached when $\bar{\langle |\Omega|\rangle}$ saturates to a constant value. This implies that the amplitude of the vorticity oscillations remains constant with $\omega\tau$ and does not significantly change with the cluster size.
Finally, the oscillation period (Fig.~\ref{fig:Fig3}~c) decreases with the reduced chirality as $1/(\omega\tau)$. This scaling confirms our intuitive explanation of this phenomenon:
Before completing a full rotation, the orientation of chiral active particles is reversed after a time period $\sim 2\pi/\omega$.
This implies that these periodic oscillations are uniquely induced by chirality.

\section{Conclusions}
The central insight of this work is that the presence of attractions in chiral active matter without alignment interactions induces self-organized vortices involving coherent dynamics of adjacent particles. These vortices can either be persistent or show a periodically oscillating vorticity, leading to patterns that self-revert their order.

The theoretical arguments developed here (for instance Eq.~\eqref{eq:4main}) could shed light on the link between chiral active systems and materials with odd properties~\cite{fruchart2023odd, tan2022odd, kole2021layered, yang2021topologically}, such as crystals characterized by odd elasticity~\cite{scheibner2020odd} and liquid governed by odd viscosity \cite{soni2019odd, lou2022odd, mecke2023simultaneous}.
Indeed, living chiral crystals exhibit self-sustained chiral oscillations as well as various unconventional deformation response behaviors recently predicted for odd elastic materials~\cite{fruchart2023odd}.
Our argument rationalizes these findings, suggesting that self-propulsion plays the role of the transverse neighbor forces typical of odd materials.

Even if here collective phenomena spontaneously emerge without alignment interactions, it could be interesting to evaluate the effects of explicit alignment mechanisms on chiral active particles at high density, in cluster configurations.
This is a rather common scenario in self-propelled colloids that can behave as chiral microswimmers by simply introducing a rotational asymmetry in their body \cite{kummel2013circular}.

This finding opens the door to the observation of customizable collective phenomena. They have the potential to inform the design and optimization of particle-based micromotors. Instead of creating asymmetric gears powered by active particles~\cite{hiratsuka2006microrotary, di2010bacterial, vizsnyiczai2017light}, spontaneous gear rotation can be achieved by harnessing chirality in active matter~\cite{li2023chirality}.
Our study could inspire experiments across a wide range of chiral active matter experiments, such as high-density chiral active colloids~\cite{kummel2013circular} attracting by means of Van-der-Waals interactions, or chiral active granular particles~\cite{scholz2018rotating, liu2020oscillating, barois2020sorting} which can be connected by springs to create crystal-like configurations~\cite{baconnier2022selective}.


\section{Methods}\label{sec:M}

\subsection{Derivation of the theoretical prediction, Eq.~(2)}\label{sec:derivation}

To derive Eq.~\eqref{eq:dynamics_explanation}, in the following we employ a similar idea as has been used in Ref.~\cite{caprini2021spatial} for straight active particles.
As we will see, accounting for chirality, leads to an additional term in the resulting equation that competes with the effective alignment that has been found in Ref.~\cite{caprini2023flocking}.
This competition is at the heart of the phenomenology which we predict and observe in the present article, as discussed in the main text.

\subsubsection{Mapping on the dynamics to an effective description}

Before proceeding to the exact mapping, it is convenient to express the dynamics of the activity in Cartesian coordinates.
By applying Ito calculus rules, Eq.~\eqref{eq:theta_dynamics} can be expressed in Ito's convention as
\begin{equation}
\label{eq:eq_n_cartesian}
\tau\dot{\mathbf{n}}_i= -\mathbf{n}_i + \omega \tau\mathbf{n}_i \times \mathbf{z} + \sqrt{2\tau}\boldsymbol{\xi}_i \times \mathbf{n}_i\,.
\end{equation}
Here, the vector $\boldsymbol{\xi}_i=(0, 0, \xi_i)$ consists only of the third component orthogonal to the plane where the particle motion takes place, namely the $xy$ plane. In this way, the noise vector can be expressed in a compact form as $\boldsymbol{\xi}_i=\mathbf{z} \xi_i$, where $\mathbf{z}$ is the unit vector normal to the $xy$ plane.

Even if the dynamics \eqref{eq:wholeABPdynamics} is underdamped, the extremely small value of the reduced inertia (i.e.\ of the inertial time compared to the persistence time) allows us to take the overdamped regime, $\dot{\mathbf{v}}_i\approx0$, so that the equation of motion for chiral active particles is effectively given by
\begin{subequations}
\label{eq:appendix_ABPcartesian_o}
\begin{align}
\label{SM:eq:v_dynamics_o}
\gamma\dot{\mathbf{x}}_i&=\mathbf{F}_i + \gamma v_0 \mathbf{n}_i + \sqrt{2T \gamma}\boldsymbol{\eta}_i \\
\label{SM:eq:theta_dynamics_o}
\tau\dot{\mathbf{n}}_i&= -\mathbf{n}_i + \tau\omega \mathbf{n}_i \times \mathbf{z} + \sqrt{2\tau}\boldsymbol{\xi}_i \times \mathbf{n}_i\,.
\end{align}
\end{subequations}
In addition, the small value of the reduced temperature (i.e.\ the small value of $T$ compared to the self-propulsion velocity square), allows us to drop the passive Brownian motion term.
By applying the time-derivative to Eq.~\eqref{SM:eq:v_dynamics_o} with $T=0$, and by defining the velocity variable, $\mathbf{v}_i=\dot{\mathbf{x}}_i$, we obtain
\begin{equation}
\gamma\dot{\mathbf{v}}_i = - \nabla^2_{x_i x_j} U_{tot} \cdot \mathbf{v}_j + \gamma v_0 \dot{\mathbf{n}}_i \,,
\end{equation}
where we have assumed Einstein's convention on repeated indices.
By replacing $\dot{\mathbf{n}}_i$ by Eq.~\eqref{SM:eq:theta_dynamics_o} immediately we have
\begin{equation}
\gamma\dot{\mathbf{v}}_i = - \nabla^2_{x_i x_j} U_{tot} \cdot \mathbf{v}_j + \gamma v_0 \left( - \frac{\mathbf{n}_i}{\tau} + \omega \mathbf{n}_i \times \mathbf{z} + \frac{\sqrt{2}}{\sqrt{\tau}} \xi_i \times \mathbf{n}_i\right) \,.
\end{equation}
Now, we proceed by replacing $\mathbf{n}$ by the Eq.~\eqref{SM:eq:v_dynamics_o} (again with $T=0$), obtaining
\begin{equation}
\begin{aligned}
\label{eq:SM_dynamicss}
\dot{\mathbf{v}}_i =& - \frac{1}{\tau}\mathbf{v}_i- \frac{\nabla^2_{x_i x_j} U_{tot}}{\gamma} \cdot \mathbf{v}_j + v_0 \frac{\sqrt{2}}{\sqrt{\tau}} \xi_i \times \mathbf{n}_i \\
&- \frac{1}{\tau}\frac{\nabla_{x_i} U_{tot}}{\gamma} +\omega \mathbf{v}_i \times \mathbf{z}
+ \frac{\omega}{\gamma } \nabla_{x_i} U_{tot} \times \mathbf{z}
\,.
\end{aligned}
\end{equation}
Dynamics~\eqref{eq:SM_dynamicss} is mathematically equivalent to Eq.~\eqref{eq:wholeABPdynamics}. In order to proceed analytically, we consider further approximations described in the next subsections.

\subsubsection{Lattice approximation for solid-like configurations}

The strong interparticle attractive interactions induce almost-perfect solid-like configurations with an almost-perfect hexagonal order.
This allows us to consider the lattice approximation by fixing the particle positions on the vertices of a triangular lattice.
In this way, every particle is characterized by six neighbors. This implies that we consider systems sufficiently large to neglect the contribution of the outer layer of particles, whose number scales as $\approx \sqrt{N_c}$. 
In this approximation, interparticle forces are perfectly balanced because of the lattice translational invariance.
As a consequence, we need only to evaluate the second derivative of the total potential in Eq. \eqref{eq:SM_dynamicss},
\begin{equation}
\begin{aligned}
\label{eq:lefthandside}
&\nabla_{x_i x_j} U_{tot} \cdot \mathbf{v}_j = \sum_{l <m}^N \nabla_{x_i x_j} U(|\mathbf{x}_l-\mathbf{x}_m|) \cdot \mathbf{v}_j\\
& =\sum_{j}^*\mathbf{v}_i \cdot\nabla_i \nabla_i U \left( r_{ij} \right) +\sum_{j}^* \mathbf{v}_j\cdot \nabla_i \nabla_j U\left( r_{ij} \right) \,.
\end{aligned}
\end{equation}
Here, $r_{ij}$ is the distance between particle $i$-th and particle $j$-th, and the sum, $\sum^*$ is restricted over the six neighbors of the target particle $i$.
The truncation at first neighbors works if the potential is short-range, as in the Lennard-Jones potential numerically considered in the numerical simulations.
To proceed further, we can calculate the spatial components of the Hessian matrix, which is a 2$\times$2 matrix, in two dimensions.
In particular, we have
\begin{equation}
\label{eq:supp_elementH}
\nabla^{\alpha}_i \nabla^{\beta}_i U\left( r_{ij} \right) = \left[ U''(r_{ij}) + \frac{U'(r_{ij})}{|r_{ij}|} \right] \frac{r_{ij}^{\alpha}r_{ij}^{\beta}}{|r_{ij}|^2} - \delta_{\alpha\beta} \frac{U'(r_{ij})}{|r_{ij}|} \,,
\end{equation}
where we have denoted the spatial components by Greek upper indices, and $r_{ij}^{\alpha} = r_i^{\alpha} - r_j^{\alpha}$, with $\alpha=x, y$. Here, each prime on the potential $U$ means a spatial derivative.
We remark that the potential depends only on the inter-particle distance and, thus the following property holds:
\begin{equation}
\nabla^{\alpha}_i \nabla^{\beta}_j U= - \nabla^{\alpha}_i \nabla^{\beta}_i U \, .
\end{equation}
To switch to a more suitable description accounting for the lattice symmetry, it is convenient to express the Cartesian components in polar coordinates, such that $r_{ij}^{x}/|r_{ij}|=\cos{\left(\delta_j\right)}$ and $\alpha=x, y$ of $r_{ij}^{y}/|r_{ij}|=\sin{\left(\delta_j\right)}$. Here, $\delta_j$ the angle between the ${\mathbf{r}}_{ij}$ vector and the $x$-axis.
\begin{widetext}
The triangular lattice structure implies the target particle $i$ has 6 first neighbors, uniquely identified by $\delta_j = \delta_0 + j\pi/3$ with $j=0, 1, ..., 5$.
The phase $\delta_0$ represents the orientation of the hexagon with respect to the reference frame that can be set to zero without loss of generality.
In this way, by denoting $|r_{ij}|=r$ (the lattice constant), we can rewrite the components of the Hessian matrix as
\begin{flalign}
\nabla^{x}_i \nabla^{x}_i U\left( r_{ij} \right)=J_{xx}(r)&=U''(\sigma)\cos^2\left(j \frac{\pi}{3}\right)+\frac{U'(\sigma)}{\sigma}\sin^2\left(j \frac{\pi}{3}\right)\\
\nabla^{y}_i \nabla^{y}_i U\left( r_{ij} \right)=J_{yy}(r)&=U''(\sigma)\sin^2\left(j \frac{\pi}{3}\right) +\frac{U'(\sigma)}{\sigma}\cos^2\left(j \frac{\pi}{3}\right) \\
\nabla^{x}_i \nabla^{y}_i U\left( r_{ij} \right)=J_{xy}(r)&=\left(U''(\sigma) - \frac{U'(\sigma)}{\sigma}\right)\cos\left(j \frac{\pi}{3}\right)\sin\left(j \frac{\pi}{3}\right)= J_{yx}(\sigma)\,.
\end{flalign}
\end{widetext}
By summarizing, the left-hand-side of Eq. \eqref{eq:lefthandside} can be expressed as
\begin{equation}
\label{SM:app:force}
\nabla_{x_i x_j} U_{\text{tot}}\cdot \mathbf{v}_j = \sum_j^* \mathbf{J}_j \left(\mathbf{v}_i - \mathbf{v}_j\right) \,,
\end{equation}
where the matrix $\mathbf{J}_j$ has elements
\begin{equation}
\mathbf{J}_j = \begin{pmatrix}
J_{xx}(\sigma)\,& J_{xy}(\sigma) \\
J_{yx}(\sigma)\,& J_{yy}(\sigma)\\
\end{pmatrix}
\end{equation}
Because of the following properties:
\begin{equation}
\begin{aligned}
\sum_j^* J_{xx}(\sigma) &= \sum_j^* J_{yy}(\sigma)\\
&= 3 \left(U''(\sigma) +\frac{U'(\sigma)}{\sigma} \right) \equiv K\noindent
\end{aligned}
\end{equation}
\begin{flalign}
&\sum_j^* J_{xy}(\sigma) = 0 \\
& \sum_j^* \mathbf{J}_j = K \boldsymbol{\mathcal{I}} \,,
\end{flalign}
we can conclude that the force \eqref{SM:app:force} has the shape of an effective alignment interaction between the particle $i$ and its 6 first neighbors.

We remark that to apply our theory the potential has to be differentiable: in particular first and second derivatives of the potential should be defined.
Our choice of Lennard Jones potential, truncated at $3 \sigma$ as usual in numerical studies, does not represent a problem for the applicability of the theory. Indeed, we resorted to the first-neighbors approximation, which allows us simply to select the interactions between the six-neighboring particles, which are at distance $\approx \sigma < 3\sigma$ where the first two derivatives of the potential are well-defined. 


\subsubsection{Effect of chirality}

\begin{figure*}[!t]
\centering
\includegraphics[width=0.9\linewidth,keepaspectratio]{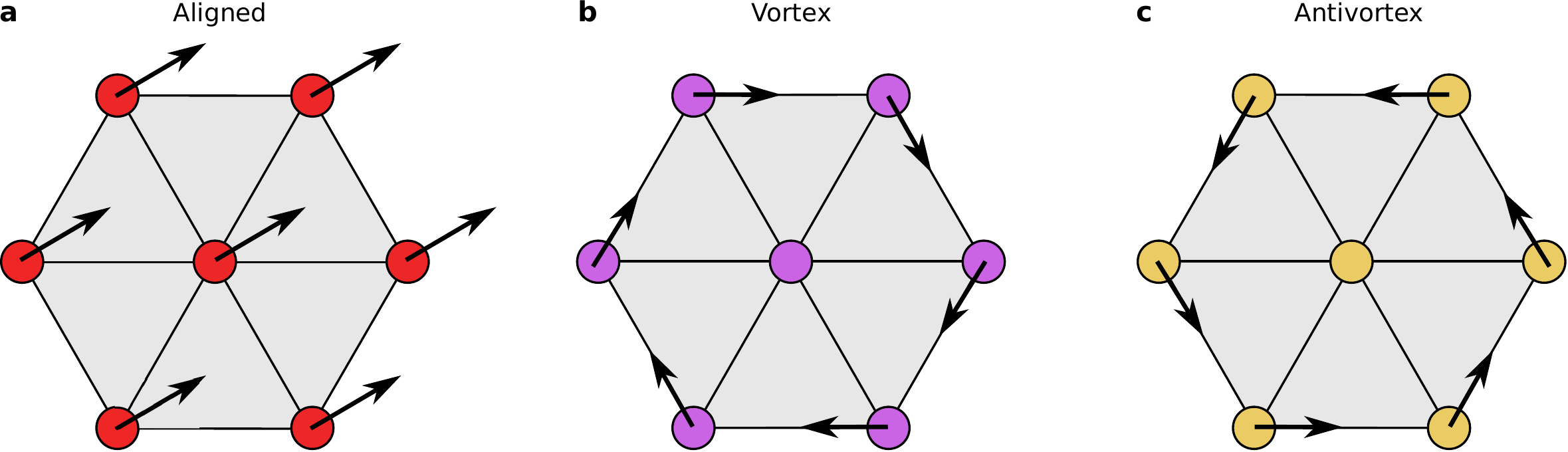}
\caption{ \textbf{Illustrations of configurations that minimize the effective alignment interactions.} The tagged particle is in the middle while neighboring particles are placed on the vertices of a hexagon. Velocities are represented by black arrows.
(a): Aligned velocities; (b) Vortex-distributed velocities; (c) Antivortex-distributed velocities.
}\label{fig:Fig1_SM}
\end{figure*}

By summarizing the results of previous sections, the behavior of chiral active particles is well-described by the following dynamics obtained after performing the lattice approximation
\begin{equation}
\label{eq:eq:app_eq2}
\dot{\mathbf{v}}_i = - \frac{1}{\tau}\mathbf{v}_i + v_0 \frac{\sqrt{2}}{\sqrt{\tau}} \xi_i \times \mathbf{n}_i - \frac{1}{\gamma}\sum_j^* \mathbf{J}_j \left(\mathbf{v}_i - \mathbf{v}_j\right) +\omega \mathbf{v}_i \times \mathbf{z} \,.
\end{equation}
Equation~\eqref{eq:eq:app_eq2} corresponds to the dynamics~\eqref{eq:dynamics_explanation}.
The first term is an effective friction force, whose friction coefficient is determined by the inverse of the persistence time $\tau$, this term here dissipates the energy injected by the thermal bath, whose amplitude is determined by $v_0/\sqrt{\tau}$.
The unconventional shape of this noise term is due to the choice of active Brownian particle dynamics which conserves the modulus of the active force and here involves the cross product with $\mathbf{n}_i$.
It is worth noting that by excluding chirality and interactions, the velocity scale is purely determined by $v_0$ while $\tau$ plays a negligible role, as expected.
Both terms are vanishing in the large persistence limit $\tau\gamma\to\infty$, becoming subleading in the dynamics.
The third term in the dynamics $- \frac{1}{\gamma}\sum_j^* \mathbf{J}_j \left(\mathbf{v}_i - \mathbf{v}_j\right) $ accounts for particle interactions and has the shape of an effective alignment interaction term spontaneously emerge from this analytical calculation.
Indeed, the particle $i$ feels a force proportional to the difference between the velocities of neighboring particles, $-(\mathbf{v}_i -\mathbf{v}_j)$, i.e. particle $i$ tends to align its velocity to those of neighboring particles.
As stated in the results, this effective alignment force is minimized in three different configurations: i) aligned velocities; ii) vortex-distributed velocities; and iii) antivortex-distributed velocities.
The three configurations are illustrated in Fig.~\ref{fig:Fig1_SM}.
In i), the tagged particle velocity $\mathbf{v}_i=\mathbf{v}$ is equal to any neighboring particle velocity (Fig.~\ref{fig:Fig1_SM}~a). Consequently, each term of the alignment force $\sum_j^* \mathbf{J}_j \cdot\left( \mathbf{v}_j - \mathbf{v}_i \right)$ independently vanishes.
In ii) and iii), the tagged particle velocity is zero while the six neighboring particle velocities are distributed on a vortex (Fig.~\ref{fig:Fig1_SM}~b) and an antivortex configuration (Fig.~\ref{fig:Fig1_SM}~c), respectively. Thus, particles on opposite vertices of the hexagon have equal velocities with opposite directions which perfectly balance.
This implies that
\begin{equation}
\sum_j^* \mathbf{J}_j \cdot\left( \mathbf{v}_j - \mathbf{v}_i \right)= \sum_j^* \mathbf{J}_j \cdot \mathbf{v}_j =0\,,
\end{equation}
or, in other words, the effective alignment interaction is not only minimized by aligned velocities but also by vortex and antivortex configurations.
Finally, the last term $\omega \mathbf{v}_i \times \mathbf{z}$ accounts for the role of chirality.
Such a force term has the shape of an effective magnetic field with amplitude $\omega$ and it is responsible at the single particle level for particle rotations.
Intuitively, this term selects vortex or antivortex configurations depending on the sign of the chirality $\omega$.

We also remark that in our theory we resort to a linearization of the force between different particles. This is possible because of the solid structure. In principle, perturbation theory can be applied to mass defects \cite{caprini2023inhomogeneous} or non-linear potentials with a weak non-linearity, such that the force
$\mathbf{F} \approx -k_0\mathbf{x}- k_{1}|\mathbf{x}|^2 \mathbf{x}$, with $k_1/k_2 \ll 1$.
In this case, we expect that the theory could quantitatively provide a correction to our results without changing the observation of the three dynamical states.

\subsection{Small persistence time regime}

In the small persistence time regime $\tau\ll\tau_I$, the cluster does not show any coherent motion and simply diffuses. Indeed, in this case, the active force $\gamma v_0\mathbf{n}$ changes fast in its direction, and it can be approximated by an effective Brownian motion. Therefore, in this regime, chirality plays a negligible role and consequently, the three states we have identified, i.e.\ non-permanent vorticity state, permanent vorticity state, and self-reverting vorticity state, cannot be observed.

This conclusion can be analytically derived by considering the dynamics \eqref{eq:v_dynamics} with active force evolving in Cartesian coordinates \eqref{eq:eq_n_cartesian}. In particular, it is convenient to express the activity dynamics by resorting to a matrix formalism
\begin{equation}
\label{eq:eq1}
\begin{aligned}
\dot{\mathbf{n}}=- \frac{\mathbf{B}}{\tau} \cdot \mathbf{n} + \frac{\sqrt{2}}{\sqrt{\tau}}\boldsymbol{\xi}
\end{aligned}
\end{equation}
where $\mathbf{B}$ is a matrix with components
\begin{equation}
\mathbf{B}=
\begin{bmatrix}
1 & \tau\omega\\
-\tau\omega & 1 \,.
\end{bmatrix}
\end{equation}
In the small persistence time regime, $\tau\ll \tau_I$, $\tau$ is the faster time scale and we can take the overdamped limit in the equation for $\mathbf{n}$, by setting $\dot{\mathbf{n}}=0$:
\begin{equation}
\label{eq:n}
\mathbf{n}=\tau \mathbf{B}^{-1}\cdot \frac{\sqrt{2}}{\sqrt{\tau}}\boldsymbol{\xi}
\end{equation}
where $\mathbf{B}^{-1}$ is the inverse of $\mathbf{B}$ with components
\begin{equation}
\mathbf{B}^{-1}=\frac{1}{1+\tau^2\omega^2}
\begin{bmatrix}
1 & -\tau\omega\\
\tau\omega & 1\,.
\end{bmatrix}
\end{equation}
By substituting Eq.\eqref{eq:n} in the dynamics \eqref{eq:v_dynamics}, we obtain
\begin{equation}
m \mathbf{v} = -\gamma \mathbf{v} + \sqrt{2 T \gamma} \boldsymbol{\eta} + \mathbf{F} + \gamma v_0 \sqrt{2\tau} \mathbf{B}^{-1}\cdot\boldsymbol{\xi} \,.
\end{equation}
As a consequence, the active force simply behaves as a white noise which cannot induce the non-equilibrium collective motion observed in the regime of large persistence time.

\subsection{Numerical methods}

\subsubsection{Dimensionless dynamics and dimensional parameters}\label{sec:1_dimensionless}

Simulations are performed by considering Eqs.~\eqref{eq:wholeABPdynamics} with rescaled variables.
Particle positions are rescaled with the particle diameter $\sigma$, so that $\mathbf{x}'=\mathbf{x}_i/\sigma$, while time is rescaled with the persistence time $\tau=1/D_r$, such that $t'=t/\tau$.
With this choice, Eqs.~\eqref{eq:wholeABPdynamics} can be integrated using the Euler method with time-step $dt' =dt/\tau=10^{-6}/\tau$ and reduces to
\begin{subequations}
\label{eq:dynamics_rescaled}
\begin{align}
&\delta \mathbf{x}_i'(t')= \mathbf{v}_i'(t') dt'\\
&\delta\mathbf{v}_i'(t') = \frac{\tau^{3/2}}{m\sigma}\sqrt{2\gamma T} \,\sqrt{dt'}\,d\boldsymbol{\eta}_i'(t')\\
&+\left( -\frac{\tau}{\tau_I} \mathbf{v}_i'(t') +\frac{\tau}{\tau_I} \frac{\tau v_0}{\sigma} \mathbf{n}_i(t') + \frac{\tau^2}{\sigma^2}\frac{\epsilon}{m} \nabla_i' U(\sigma\mathbf{x}'(t')) \right) dt'\nonumber\\
&\delta\theta_i(t') =\sqrt{2}\,\sqrt{dt'}\, d\xi_i(t') +\tau\omega dt'\,,
\end{align}
\end{subequations}
where $\delta \mathbf{x}'_i(t) =\mathbf{x}_i'(t+dt)-\mathbf{x}_i'(t)$ and $\delta \mathbf{v}'_i(t) =\mathbf{v}_i'(t+dt)-\mathbf{v}_i'(t)$ are the increment of particle position and velocity after a time-step $dt'$, while $\delta \mathbf{\theta}'_i(t) =\mathbf{\theta}_i'(t+dt)-\mathbf{\theta}_i'(t)$ represents the time integral of the orientational angle of the particle $i$.
In addition, $d\boldsymbol{\eta}_i'(t')$ and $d\xi_i(t')$ are two dimensionless Wiener processes with zero average that can be numerically generated by Gaussian numbers with unit variance and we have used the definition of the inertial time $\tau_I=m/\gamma$.
The dynamics \eqref{eq:dynamics_rescaled} is governed by five dimensionless parameters that are listed and commented on below:
\begin{enumerate}[(i)]
\item Reduced inertial time $\tau_I/\tau=m/(\gamma \tau)=10^{-6}$ which determines the velocity relaxation in units of persistence time.
\item P\'eclet number $\text{Pe}=\tau v_0/\sigma=50$, which quantifies the power of the active force.
\item Reduced energy scale $\tau^2 \epsilon/(\sigma^2 m)=5\times 10^{3}$, which determines the strength of the attractive force.
\item Reduced temperature $\tau^2 T /(m\sigma^2)=10^{-5}$, which quantifies the effect of the thermal noise on the system.
\item Reduced chirality $\omega\tau$, which governs the time-scale associated with chirality and it is varied in the simulations to address its effect.
\end{enumerate}
With this choice of parameters in particular $\tau_I/\tau=10^{-6}$, the dynamics~\eqref{eq:wholeABPdynamics} (or the dimensionless Eqs.~\eqref{eq:dynamics_rescaled}) are effectively in the overdamped regime.
However, since Eqs.~\eqref{eq:dynamics_rescaled} is an underdamped equation of motion, velocities $\mathbf{v}_i$ are well-defined.
The underdamped choice is particularly convenient to calculate velocity and vorticity fields because they remain well-defined even in the presence of thermal noise. 
However, the numerical results reported in this paper can be also observed with an overdamped active model if the thermal noise is sufficiently small.
In addition, in numerical simulations, we explore different cluster size square $\sigma^2 N= 113, 226, 452, 904, 1809$ and packing fraction $\phi= N/L^2 \pi \sigma^2 =0.35$. The size of the box $L$ is chosen accordingly.
The system spontaneously evolves to a state characterized by a unique cluster because of attractive interactions. However, depending on the total number of particles in simulations, the system could take a long transient time to reach the steady state. Thus, when needed, simulations were directly initialized in the cluster configuration.

\subsubsection{Details on the distinction beween the different states}

In Fig.~\ref{fig:Fig2}~a, we have distinguished between three states:
\begin{enumerate}[i)]
\item Non-permanent states (black dots in Fig.~\ref{fig:Fig2}~a). This state is characterized by fluctuating vorticity, from negative to positive values, which is compatible with configurations with negligible chirality.
\item Permanent vorticity states (orange dots in Fig.~\ref{fig:Fig2}~a). In this state, the cluster is characterized by a permanent vorticity and displays a permanent rotating trajectory aligned to the particle chirality.
\item Vortex-antivortex state (violet dots in Fig.~\ref{fig:Fig2}~a). This state shows the self-reverting vorticity observed and the cluster spinning dynamics.
\end{enumerate}
States i), ii), and iii) are characterized by a continuous crossover rather than a sharp phase transition. This feature is already evident by evaluating the time-averaged value of the spatial average vorticity $\bar{\langle \Omega\rangle}\tau$ (Fig.~\ref{fig:Fig3}~a)
and the time-average of the modulus of the rescaled spatial average vorticity $\bar{\langle |\Omega|\rangle}\tau$ (Fig.~\ref{fig:Fig3}~b.
In particular, the first two states are distinguished by comparing the time fluctuations and time average of the total vorticity field $\langle \Omega(t)\rangle$.
In particular, configurations that belong to state i) are characterized by a time-standard deviation of $\langle \Omega(t)\rangle$ larger than its average, i.e. by the following relation:
\begin{equation}
\label{eq:average_flu_cond}
\frac{1}{t}\int dt \langle \Omega(t)\rangle < \left( \frac{1}{t}\int dt \langle \Omega(t)\rangle^2 \right)^{1/2} \,,
\end{equation}
holding for $t\to\infty$.
By contrast, configurations that belong to state ii) satisfies
\begin{equation}
\frac{1}{t}\int dt \langle \Omega(t)\rangle > \left( \frac{1}{t}\int dt \langle \Omega(t)\rangle^2 \right)^{1/2} \,,
\end{equation}
for $t\to\infty$.
Finally, configurations belonging to state iii) again satisfy the condition Eq.~\eqref{eq:average_flu_cond}.
However, at variance with state 1, $\langle \Omega(t)\rangle$ switches from negative to positive values periodically in time.

\subsection{Derivation of the theoretical argument (4) }\label{sec:anal}

To calculate the total torque on the cluster, let us consider the total force exerted by each microscopic active particle, $\boldsymbol{\mathcal{F}}_i$, given by the sum of attractive interactions and active forces
\begin{equation}
\boldsymbol{\mathcal{F}}_i = - \sum_j^* \nabla_i U(|\mathbf{x}_i -\mathbf{x}_j|) + \gamma v_0 \mathbf{n}_i \,.
\end{equation}
where the sum $\sum_j^*$ runs over the six neighbors of the $i$-th particle.
To calculate the torque due to the particle $i$-th, we have to apply the vector product of the relative particle position calculated from the center of the cluster $\mathbf{r}_i$
\begin{equation}
\mathbf{T}_i = \mathbf{r}_i\times\boldsymbol{\mathcal{F}}_i \,.
\end{equation}
By summing over $i$, the contribution of the internal force vanishes by symmetry and the total torque reads
\begin{equation}
\mathbf{M}=\sum_{i=0}^{N_c} \mathbf{T}_i = \gamma v_0 \sum_{i=0}^{N_c} \mathbf{r}_i\times \mathbf{n}_i\,.
\end{equation}
By assuming that clusters have spherical shapes, the torque can be decomposed as
\begin{equation}
\mathbf{M}=\int_0^R dr' \mathbf{m}(r')
\end{equation}
where $r'$ is the radial coordinate with respect to the center of the cluster and $R$ is the cluster radius.
As a consequence, $\mathbf{m}(r')$ is the torque due to the particles at distance $r'$ from the cluster center and can be expressed as
\begin{equation}
\mathbf{m}(r')=\sum_{i=0}^{N_c} \mathbf{T}_i \,\delta(|\mathbf{r}_i|-r') = \gamma v_0 \sum_{i=0}^{N_c} \mathbf{r}_i\times \mathbf{n}_i \,\delta(|\mathbf{r}_i|-r')\,.
\end{equation}
This expression corresponds to Eq.~\ref{eq:4main} after recognizing that, as a first approximation, $\mathbf{M} \approx \mathbf{m}(R)$, since the particles in the outer layer provide the larger contribution to the torque being those at the larger distance from the center.

\subsection{Description of the Supplementary Movies}\label{sec:SupplementaryMovies}

Supplementary Movie 1 and Supplementary Movie 2 report the time evolution of the system from two simulations in the steady-state, for reduced chirality $\omega\tau=10$ and $2\times10^2$, respectively. The first shows a typical configuration in the permanent vorticity state, while the second displays a typical one in the self-reverting vorticity state.
In both cases, the cluster size square is given by $L_c^2=904$, corresponding to a number of particles in the cluster, $N_c=904$.
Each movie consists of two adjacent videos lasting for a total time interval $10^2\tau$.
In the left video, particles are plotted as circular points and colored according to their orientational angle $\theta_i$.
In the right video, we plot the coarse-grained fields: black arrows are used to denote the velocity field $v(\mathbf{r})$, while the color gradient shows the vorticity field $\Omega(\mathbf{r})$.
In both cases, orientations $\theta_i$ are random since there are no torques between different particles, and in general, the velocity field $v(\mathbf{r})$ shows global alignment characterized by vortex-like configurations.
The main differences between the two movies appear in the global cluster motion and in the time evolution of the vorticity field $\Omega(\mathbf{r})$.
In Supplementary Movie 1, the cluster shows a global counterclockwise rotation and $\Omega(\mathbf{r})$ is locally and globally always larger than zero.
In Supplementary Movie 2, the cluster is characterized by spinning dynamics while $\Omega(\mathbf{r})$ displays local oscillations between negative and positive values.

\section{Data availability}
The data that support the plots within this paper and other findings of this study are available from the corresponding author upon request, while Supplementary Movie 1 and Supplementary Movie 2 are uploaded as Supplemental Material.

\section{Code availability}
The code to generate data, by using numerical simulations, is available under reasonable requests.

\bibliographystyle{naturemag}
\bibliography{bib}

\section{Acknowledgments} 
We thank Jens Elgeti and Benoit Mahault for helpful discussions.
LC acknowledges support from the Alexander von Humboldt foundation and acknowledges the European Union MSCA-IF
fellowship for funding the project CHIAGRAM.
HL acknowledges support by the Deutsche Forschungsgemeinschaft (DFG) through the SPP 2265 under the grant number LO 418/25.

\section{Author contributions}
LC performed the numerical study and derived the theoretical predictions.
LC, BL, and HL equally contributed to the data interpretation, presentation, and writing of the manuscript.

\section{Competing Interests}
The authors declare no competing interests.

\end{document}